\documentclass[preprint,aps,prl,showpacs]{revtex4}
\usepackage{epsfig}
\usepackage{amsfonts}
\begin{document}

\title{Globally-Linked Vortex Clusters in Trapped Wave Fields}

\author{Lucian-Cornel Crasovan}

\altaffiliation[Permanent address:]{Institute of Atomic Physics,
Bucharest, Romania.}

\author{Gabriel Molina-Terriza}
\author{Juan P. Torres}
\author{Lluis Torner}

\affiliation{Laboratory of Photonics, Universitat Politecnica de
Catalunya, 08034 Barcelona, Spain}

\author{V\'{\i}ctor M. P\'erez-Garc\'{\i}a}

\affiliation{Departamento de Matem\'aticas, E.T.S.I.~Industriales,
  Universidad de Castilla-La Mancha, 13071 Ciudad Real,
Spain}

\author{Dumitru Mihalache}

\affiliation{Department of Theoretical Physics, Institute of
Atomic Physics, Bucharest, RO 76900, Romania}

\begin{abstract}
We put forward the existence of a rich variety of fully stationary
vortex structures, termed H-clusters, made of an increasing number
of vortices nested in paraxial wave fields confined by trapping
potentials. However, we show that the constituent vortices are
{\it globally linked}, rather than products of independent
vortices. Also, they always feature a {\it monopolar} global wave
front and exist in nonlinear systems, such as Bose-Einstein
condensates. Clusters with multipolar global wave fronts are
non-stationary or at best flipping.
\end{abstract}

\pacs{42.25.-p} \maketitle

Singular wave structures, that contain topological wave front
dislocations \cite{NyeBer74}, are ubiquitous in many branches of
classical and quantum science.  Screw dislocations, or vortices,
are a common dislocation type.
Wave packets with nested vortices find applications in fields as
diverse as cosmology, biosciences, or solid state physics
\cite{Lug95,Pis99,Ash99,GalOrm01}.  As striking examples, they are
at the heart of schemes to generate engineered quNits in quantum
information systems in higher dimensional Hilbert spaces
\cite{MaiVazWei01,MolTorTor02}, are believed to be essential for
the onset of superfluidity in Bose-Einstein condensates (BEC)
\cite{Don91,BEC-exp,Butts,Tsubota}, or allow tracking the motion
of a single atom \cite{Hor02}.

A recent study of the motion of vortex lines governed by both
linear and nonlinear Scr\"odinger equations describing the
dynamics of atoms in harmonic traps revealed that the topological
features of vortex dynamics are to large extent universal
\cite{polish}. The dynamics of the vortices nested on localized
wave packets  depends on the evolution of the host beam, and on
the interferences and interactions between the vortices
\cite{reviewsOV}.  Multiple vortices nested on the same host
typically follow dynamical evolutions which might include large
vortex-drifts that destroy their initial arrangement, and
vortex-pair annihilations that destroy the vortices themselves.
Vortex evolutions are particularly complex in strongly nonlinear
media, such as BEC, where the vortices can interact with each
other. Therefore, a fundamental question arises about whether
stationary or quasi-stationary vortex clusters or lattices
\cite{latticesinBEC}, made of vortices with equal and with
opposite topological charges exist. To isolate the pure vortex
features from the dynamics solely induced by the evolution of the
host wave packet, it is convenient to study wave fields confined
by suitable potentials, as in weakly-interacting trapped BEC.

In this paper we show that vortex clusters with multipolar global
wave fronts nested in wave fields confined by trapping potentials
are non-stationary, when the number of vortices and their location
are not constant during dynamical evolution, or at best flipping,
when the vortices periodically flip their topological charges
through extremely sharp Berry trajectories \cite{Ber98}. In the
former case, multiple vortex revivals mediated by Freund
stationary point bundles \cite{Fre00}, that carry the necessary
Poincar\'e-Hopf indices \cite{Fre01}, can occur. In contrast, we
find that a rich variety of {\it fully stationary\/} vortex
clusters made of an increasing number of vortices do exist. The
important point we put forward is that these clusters are {\it
globally linked}, rather than products of independent vortices.
Also, they feature a {\it monopolar} global wave front.  We also
show that the clusters exist and are robust in nonlinear systems
such as interacting BEC.

We thus address the slowly-varying evolution of generic
wavefunctions governed by the paraxial wave equation
\begin{equation}
iA_{z} = {\cal L} A + {\cal{N}} (A),
 \label{ecgen}
\end{equation}
where $A$ is a complex field, ${\cal L}$ is a two-dimensional
linear differential operator containing a trapping potential and
${\cal{N}}(A)$ takes care of any nonlinear contribution. We assume
the trapping potential to be harmonic thus ${\cal L}=-1/2\,\left(
\partial^2_x + \partial^2_y \right)+  \left( n_x x^2+n_y y^2
\right).$ To be specific, when ${\cal{N}}(A)\sim |A|^2A$, this
equation models the propagation of a light beam guided in a Kerr
nonlinear graded-index medium and the mean-field evolution of a
two-dimensional trapped BEC at zero-temperature (where $n_{x,y}$
are proportional to the trap frequencies in appropriate units).
Here we will consider only the symmetric case; hence $n_x=n_y=2$.
For convenience, from now on we will split
$A(x,y;z)=F(x,y;z)V(x,y;z)$, taking the host packet $F$ as given
by the fundamental mode of the trapping potential
$F(x,y;z)=\exp(-x^2-y^2) \exp(-2iz)$. In the linear case, the
function $V(x,y;z)$ carries all the essential information about
the solutions and in particular about vortex dynamics. Here we
will consider the evolution of polynomial initial data for $V$
corresponding to (multi)vortex solutions of Eq.~(\ref{ecgen}).
Such solutions can be expressed as finite series and, as will be
clear later, all of them must be periodic or stationary.

Let us first consider the linear ($\mathcal{N}(A) = 0$) evolution
of vortex-clusters built as products of $n$ independent
single-charge vortices: $V(x,y; z=0)=\prod_{k=1}^{n}
[x-x_k+i\sigma_k(y-y_k)]$, where $(x_k, y_k)$ are the locations in
the vortex cores in the transverse plane, and $\sigma_k=\pm 1$.
None of the above product-vortex clusters is found to be
dynamically stationary. On the contrary, the number of vortices
and their location is found to vary during evolution, so that the
initial vortex structure is destroyed. These results can be
illustrated by examining the evolution of the 4-vortex cluster:
$V(z=0)=(x+a+iy)(x-a+iy)(x-iy-ia)(x-iy+ia)$, which contains two
vortices with positive topological charge and two vortices with
negative charge in a symmetrical geometry.  This cluster features
a quadrupolar global wave front, as is revealed by calculating the
gradient of the wave front $\Phi$ far from the cluster core, to
obtain $|\nabla \Phi| \sim 1/\rho^3$, where $\rho$ is the polar
coordinate, similar to the corresponding electrostatic multipole
\cite{multipoles}.  Substitution into (\ref{ecgen}) yields
\begin{eqnarray}
V(x,y; z)& = & [(x^2+y^2)(x^2+y^2+2 e^{4iz}-2)]e^{-8iz}\nonumber \\
& & + 4ia^2xye^{-4iz}+1/2(1-e^{-4iz})^2-a^4. \label{solc1}
\end{eqnarray}
 One finds three different regimes of evolution, as shown in
Fig.~1: Vortex drifts, and vortex-pair annihilations and revivals
take place, so that depending on the value of the geometrical
parameter $a$, the total number of vortices, $n$, hosted in the
wave field during propagation can oscillate between: (i) 4 and 8
[see Fig.~1(a)]; (ii) 4, 0 and 8 [see Fig.~1(b)]; (iii) 4 and 0
[see Fig.~1(c)].  When $n=0$, it is understood that all vortices
have annihilated each other. Analogous evolutions were found for
octupoles and higher-order multipoles.  Only dipoles can be made
quasi-stationary, but flipping, when the corresponding vortex
twins periodically flip their topological charges. Thus, the main
conclusion reached is that the interference between the
constituent vortices of all the product-clusters produces beatings
between the normal modes of the system, rendering the clusters
non-stationary.


The key insight we put forward in this paper is that such beatings
are not associated to the intrinsic or local properties of the
individual vortices, but to the very way the vortices are {\it
globally linked\/} in the host wave packet. As an example, let us
consider the evolution of
\begin{equation}
\label{4vcl}
 V(x,y; z=0)=x^2+y^2-a^2+2ixy,
  \end{equation}
 which contains four
vortices located at the same positions and having the same charges
as those of the vortex-quadrupole considered above [see Fig.~2].
However, in this cluster the vortices are intimately linked to
each other, rather than individually nested in the host $F$. This
fact manifests itself in the global wave front of the cluster,
which behaves as $|\nabla \Phi| \sim \cos(2\phi)/\rho$ +
${\cal{O}}(1/\rho^3)$, and thus features a {\it monopolar} decay
almost everywhere.  In this case, the vortex evolution is given by
\begin{equation}
\label{4vcl_evol}
 V(x,y;z)=(x^2+y^2-1/2+2ixy)e^{-4iz}+1/2-a^2,
  \end{equation}
an evolution that when the cluster is constructed with
$a=1/\sqrt{2}$ does become fully stationary.


The above 4-vortex cluster is not an isolated case, but rather an
example of whole existing families of fully stationary vortex
structures made of globally linked vortices.  In fact, the
solutions of Eq.~(\ref{ecgen}) with  ${\cal{N}} (A)=0$ have the
form:
\begin{equation}
\label{herm0} A(x,y;z) = \sum_{k,l=0}^{\infty} C_{kl}
H_{k}(x\sqrt{2})H_{l}(y\sqrt{2}) e^{-x^2-y^2}e^{-2i(k+l+1)z},
\end{equation}
where $H_j$ are the Hermite polynomials. Therefore, the evolution
of initial data of the form $V(x,y; z=0) = \sum_{k=0}^{n}C_{k}
H_k(\xi) H_{n-k}(\eta)$ for any $C_{k} \in \mathbb{C}$ and $\xi =
x\sqrt{2}, \eta = y\sqrt{2}$, is given by
\begin{equation}
\label{Hermite} V(x,y;z) = e^{-2inz} \sum_{k=0}^{n}C_{k} H_k(\xi)
H_{n-k}(\eta).
\end{equation}
On physical grounds, this simple mathematical result shows that
all the stationary clusters are made of globally-linked vortices.
Equation (\ref{Hermite}) allows us to build a variety of
structures, to be termed Hermite, or H-clusters, whose key
features we discuss in what follows.

Perhaps the simplest type of H-clusters are those with a $n\times
n$ {\it matrix} geometry thus containing $n^2$ vortices. These
clusters can be generated by using as initial data, for example,
$V_{n\times n }(x,y; z=0)=H_n(\xi)+iH_n(\eta)$. In this function
the vortex charges alternate throughout the matrix and the vortex
locations are dictated by the zeroes of the particular Hermite
polynomials involved. In general these vortex matrices are not
regular, the distance between vortices varying along the matrix.
However, in the particular cases with $n=2$ and $n=3$ the matrix
is regular. The $n=3$ case is shown in Fig.~2(a). Notice that a $2
\times 2$ matrix cluster can be generated either with $V(x,y;
z=0)=H_2(\xi)+iH_2(\eta)$, or with $V(x,y;
z=0)=H_2(\xi)+H_2(\eta)+iH_1(\xi)H_1(\eta)$. Actually, this latter
possibility generates the stationary 4-vortex cluster discussed
earlier [see Fig.~2(b)].

One can also build $m\times n$ ($m \ne n$) stationary vortex
matrices. A possible choice for the vortex function generating
such a vortex-matrix is $V_{m\times n}(x,y;
z=0)=H_m(\xi)+iH_{|m-n|}(\xi)H_n(\eta)$. As an example we show the
$4 \times 2$ vortex matrix in Fig.~2(c). In contrast to the $n
\times n$ matrices, in the general case the topological charges
carried by the vortices of the $m\times n$ matrices do not
alternate sign throughout the matrix.  An important subclass of
the $m\times n$ vortex matrices are the $m\times 1$ cases, to be
termed {\it vortex arrays}. They consist in $m$ co-linearly
displaced vortices of the same topological charge. Figures 2(d)
and 2(e) show illustrative examples. The simplest array is the
vortex-twin shown in Fig.~2(d): A pair of identical vortices that,
contrary to the vortex dipole which either undergoes periodic
annihilations and revivals or charge flip-flops, can be made fully
stationary.

The $n\times n$ matrices are either chargeless for even $n$, or
carry a single net charge for odd values of $n$, while the
$m\times 1$ arrays carry a $m$ total topological charge. In any
case, the wave front of all the H-clusters is found {\it to
feature a monopolar} decay ($\sim 1/\rho$) almost everywhere.

More complex H-clusters also exist, and a full classification of
all the possibilities falls beyond the scope of this paper.
However, an example of one of such exotic H-clusters is displayed
in Fig.~2(f), which corresponds to the cluster built with $V(x,y;
z=0)=H_3(\eta)+i(H_3(\xi)+H_1(\eta)H_2(\xi))$. The rich variety of
possibilities  contained in Eq.~(\ref{Hermite}) is clearly
apparent.


An interesting issue is the existence and stability of
vortex-clusters in the presence of nonlinear cubic interactions
such as those appearing in the propagation of beams in Kerr media
or in the dynamics of BEC. To ease the comparison with BEC
literature we choose now $n_x = n_y = 1/2$ and $\mathcal{N}(A) = U
|A|^2 A$ \cite{nota}. In this context the evolution variable is
denoted by $t$ instead of $z$. With this choice of parameters the
range of $U$ values experimentally accesible for the
two-dimensional case is $0 < U < 10^2-10^3$ \cite{GRPG}.


We have studied several particular examples to verify that these
structures indeed exist and are stable in the nonlinear regime. We
have taken as initial data several linear configurations such as a
single vortex, dipole systems and the 4-vortex cluster given by
Eq. (\ref{4vcl}) and evolved them for $UN = 10$ ($N =\int |A|^2
d\mathbf{x}$ is the wave function norm) using an standard
split-step integrator. It is found that, although the background
performs oscillations and the vortex locations oscillate around
their equilibrium positions, the vortex clusters remain stable
(see Fig.~3). We have also searched for stationary solutions of
Eq. (\ref{ecgen}), of the form $A(x,y;t) = e^{i\lambda t}
\psi(x,y)$. To do so we have used a steepest descent method to
minimize the functional \cite{SIAM}:
\begin{equation}
\label{funct} F(\psi) = \frac{\int \psi^* \left(-\Delta -\lambda +
r^2 + U|\psi|^2\right) \psi d\mathbf{x}}{\int |\psi|^2
d\mathbf{x}},
\end{equation}
whose minima (except for $\psi = 0$) coincide with the stationary
solutions of Eq. (\ref{ecgen}) for a given value of $\lambda$.
 For instance, taking as initial data for the minimization process
 the linear 4-vortex cluster and setting $\lambda = 8.0$ and $U =
 100$, we found a stationary 4-vortex cluster solution (see Fig.~4)
 with norm $N  =
\int |\psi|^2 d\mathbf{x} \simeq 1.6005$ (thus the product $UN
\simeq 160$ which lies into the fully nonlinear regime). We have
verified that this solution is robust under time evolution when
small perturbations are added. These evidences
 show that the existence of H-clusters in a BEC should be
experimentally accessible, at least from the dynamical point of
view.


To conclude, we stress that the constituent vortices of the
H-cluster are {\it globally linked}, rather than products of
independent vortices. Following this idea it is possible to
generate a variety of additional novel structures with fascinating
properties. A nice new example are the circular vortex necklaces
generated at the intersection between the circle $x^2+y^2-a^2=0$,
where $\Re e(V)=0$, and the lines $y \pm \tan(2k\pi/n)x=0$, $k \in
\mathbb{N}$, where $\Im m(V)=0$. Those are quasi-stationary,
purely flipping clusters made of $n$ vortices (see Fig.~5), a
feature so far only known to occur with vortex-dipoles. Once
again, the vortices forming the necklace are intimately linked and
do not exhibit a $n$-polar wave front. The exploitation of such
intrinsic linking might open new opportunities in classical and
quantum systems based on topological light and matter waves. The
first challenge is the demonstration of the generation of the
clusters, by suitable computer-generated holograms \cite{masks} in
Optics and phase-imprinting techniques in BEC
\cite{phase-imprinting}.

This work was supported by the Generalitat de Catalunya and
Ministerio de Ciencia y Tecnolog\'{\i}a under grants TIC2000-1010
and BFM2000-0521. L. -C. Crasovan acknowledges NATO support and
thanks M. Damian and O. Halmaghi for helpful discussions.

\newpage

\normalsize
\noindent { \bf Figure captions }

\noindent {\bf Fig. 1}: Evolution of a vortex quadrupole,
constructed as the product of four single-charge vortices. Upper
row: Number of vortices $n$ as a function of $z$, for different
values of the initial quadrupole size. (a) $a=0.5$, (b) $a=1.1$
and (c) $a=1.2$. Bottom row: Intensity snapshots corresponding to
points labeled A, B and C in (c). Black-filled circles: positive
vortices; white-filled circles: negative vortices.

\vspace{0.2cm}

\noindent {\bf Fig. 2}: H-cluster zoology. Shown are several
examples of stationary vortex matrices and vortex arrays that can
be constructed. See text for details. Lines show zero crossings of
$\Re e(V)$ (full lines) and $\Im m(V)$ (dashed). Features as in
Fig.~1.

\vspace{0.2cm}

\noindent {\bf Fig. 3}: Stable evolution of initial data given by
Eq. (\ref{4vcl}) for $UN=10$. Upper row: Intensity plots; bottom
row: Interference fringes. Spatial region spanned is
$[-4,4]\times[-4,4]$.

\vspace{0.2cm}

\noindent {\bf Fig. 4}: Linear 4-vortex cluster [Eq. (\ref{4vcl})]
vs. its nonlinear stationary version for $U=100, \lambda = 8$. (a)
Plots of $|\psi(x,y=0)|^2$ for the linear (dashed line) and
nonlinear (solid line) cases. (b,c) Surface plots of
$|\psi(x,y)|^2$ for (b) the linear and (c) nonlinear situations.
The vortex locations and topological charges are indicated by plus
and minus signs.

\vspace{0.2cm}

\noindent {\bf Fig. 5}: Evolution of a flipping $n=8$ circular
vortex-necklace. Vortex pattern at (a) $z=0$ and (b) $z=3\pi/16$.


\begin{thebibliography}{200}
\bibitem{nota} With this choice  the basis
functions are $H_m(x), H_n(y)$ and the ``background" $F(x,y,t) =
e^{-x^2/2-y^2/2}e^{it/2}$.
\end{thebibliography}
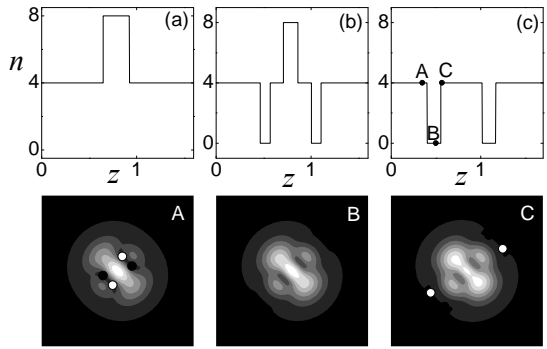
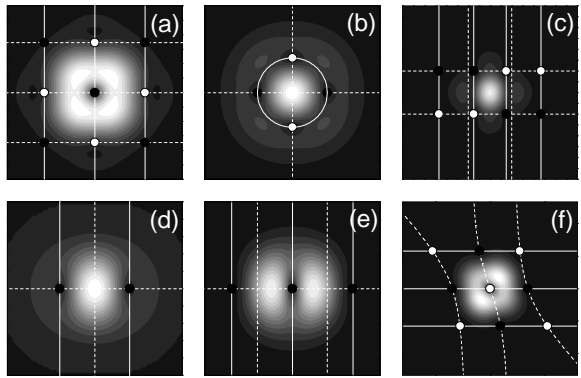
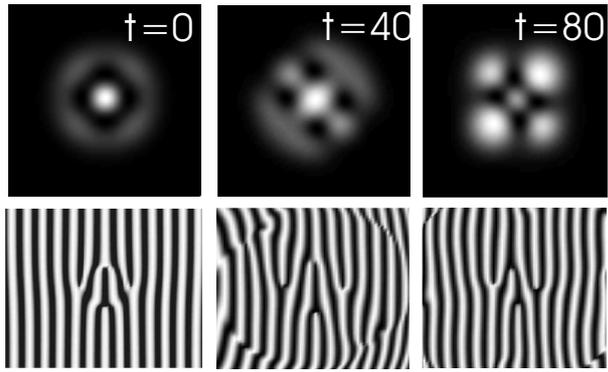
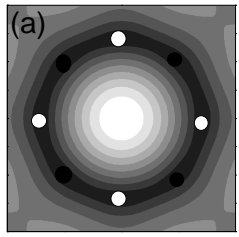
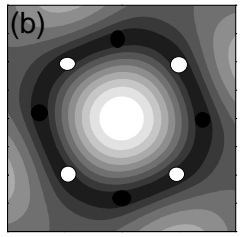
\end{document}